\documentclass[manuscript]{aastex}

\newcommand\gsim{\,\lower3pt\hbox{$\sim$}\llap{\raise2pt\hbox{$>$}}\,}
\newcommand\lsim{\,\lower3pt\hbox{$\sim$}\llap{\raise2pt\hbox{$<$}}\,}

\usepackage{emulateapj5}
\usepackage{graphicx}
\usepackage{epsfig}
\usepackage{subfigure}
\usepackage{natbib}

\shortauthors{LUGAZ ET AL.}
\shorttitle{AZIMUTHAL PROPERTIES OF CMES}

\authoraddr{N.\ Lugaz,  I.~I.\  Roussev
Institute for Astronomy, University of Hawaii
2680 Woodlawn Drive, Honolulu, HI 96822, USA.
(nlugaz@ifa.hawaii.edu)}

\authoraddr{J.~N.\ Hernandez-Charpak,
Universidad de los Andes, Bogota, Colombia.}

\authoraddr{C.~J. Davis, J.~A. Davies
STFC Rutherford Appleton Laboratory,  Chilton, Didcot, Oxfordshire, OX11 0QX, UK.}

\authoraddr{A. Vourlidas,
Code 7663, Naval Research Laboratory, Washington, DC 20375, USA.}

\begin{document}


%
%

\title{DETERMINING THE AZIMUTHAL PROPERTIES OF CORONAL MASS EJECTIONS FROM MULTI-SPACECRAFT REMOTE-SENSING OBSERVATIONS WITH {\it STEREO} SECCHI}
\author{N.\ Lugaz \altaffilmark{1}, J.~N.\ Hernandez-Charpak\altaffilmark{2}, I.~I.\ Roussev \altaffilmark{1}, C. J. Davis \altaffilmark{3}, A. Vourlidas \altaffilmark{4}, J.~A. Davies \altaffilmark{3}}
\altaffiltext{1}{Institute for Astronomy, University of Hawaii, Honolulu, HI, USA; nlugaz@ifa.hawaii.edu, iroussev@ifa.hawaii.edu}
\altaffiltext{2}{Universidad de los Andes, Bogota, Colombia; jorge-he@uniandes.edu.co}
\altaffiltext{3}{STFC Rutherford Appleton Laboratory, Didcot, UK; chris.davis@stfc.ac.uk, jackie.davies@stfc.ac.uk}
\altaffiltext{4}{Naval Research Laboratory, DC, USA}

%
%

\begin{abstract}

We discuss how simultaneous observations by multiple heliospheric imagers can provide some important information about the azimuthal properties of Coronal Mass Ejections (CMEs) in the heliosphere.
We propose two simple models of CME geometry that can be used to derive information about the azimuthal deflection and the
azimuthal expansion of CMEs from SECCHI/HI observations. We apply these two models to four CMEs well-observed by both {\it STEREO} spacecraft  during the year 2008. We find that in three cases, the joint {\it STEREO-A} and {\it B} observations are consistent with CMEs moving radially outward. In some cases, we are able to derive the azimuthal cross-section of the CME fronts, and we are able to measure the deviation from self-similar evolution. The results from this analysis show the importance of having multiple satellites dedicated to space weather forecasting, for example in orbits at the Lagrangian L4 and L5 points.
 
\end{abstract}
\keywords{scattering --- MHD --- Sun: corona --- Sun: coronal mass ejections (CMEs)}

\section{Introduction} \label{intro}

With the launch of the Solar Mass Ejection Imager (SMEI) and the {\it Solar Terrestrial
Relations Observatory (STEREO)} in 2003 and 2006, respectively, Coronal Mass Ejections (CMEs) can now be observed remotely all the way to the Earth \citep[]{Harrison:2009}, 
and their properties compared with in-situ observations at 1~AU \citep[]{Davis:2009, Rouillard:2009c}. Although the Heliospheric Imagers \citep[HI-1 and HI-2, see][]{Eyles:2009}, 
part of the Sun Earth Connection Coronal and Heliospheric Investigation (SECCHI) suite
onboard {\it STEREO} \citep[]{Howard:2002}, enable the tracking of CMEs, they cannot provide directly the CME position because what is observed is the Thomson scattered signal integrated over
a line-of-sight. Therefore, the CME position and kinematics have to be derived from a variety of models, which make assumptions regarding the CME shape and its direction of propagation
\citep[]{Vourlidas:2006, Sheeley:1999, Tappin:2009}. In \citet{Lugaz:2009b}, using real and simulated data from \citet{Lugaz:2009a}, we compared some of the assumptions typically used: i) Fixed-$\phi$, where it is assumed that the HIs track a single plasma element moving radially outward, ii) Point-P, where the CME is treated as an expanding sphere centered on the Sun, and iii) a newly proposed harmonic mean (HM) model, where the CME cross-section is a circle anchored at the Sun and its diameter corresponds to the harmonic mean of the positions derived by the Fixed-$\phi$ and Point-P approximations. We found that the HM model gives the best results for a wide CME observed by one spacecraft on the limb, while the Fixed-$\phi$ approximation, which is the most widely used, gives comparable results up to approximatively 0.5~AU.

Most of the work so far on SECCHI/HI data has focused on observations by a single spacecraft, whereas stereoscopic CME observations by the SECCHI coronagraphs (COR-1 and COR-2) have resulted in a number of studies to improve our understanding of CME physical parameters. For example, \citet{Thernisien:2009} used multi-spacecraft observations to determine the CME direction, speed and orientation, and \citet{Colaninno:2009} used multi-spacecraft observations to determine the CME direction and mass. In \citet{Lugaz:2005}, we showed how CME images made by wide-angle white-light imagers from different viewpoints can help in determining the CME direction of propagation in the heliosphere. Recently, \citet{Wood:2009} proposed a technique similar to that of \citet{Thernisien:2009} but which is expanded to the HI fields-of-view and with a full treatment of the Thomson scattering. The authors use a geometrical model, which is fitted visually to joint observations by the two {\it STEREO} spacecraft to derive the CME aspect, density structure and orientation. \citet{Liu:2010}, in a recent study, calculated the CME central longitude by triangulation in the COR and HI field-of-views for one CME event and obtained good results up to approximatively 0.5~AU.

\begin{figure*}[ht*]
\begin{center}
{\includegraphics*[width=8.cm]{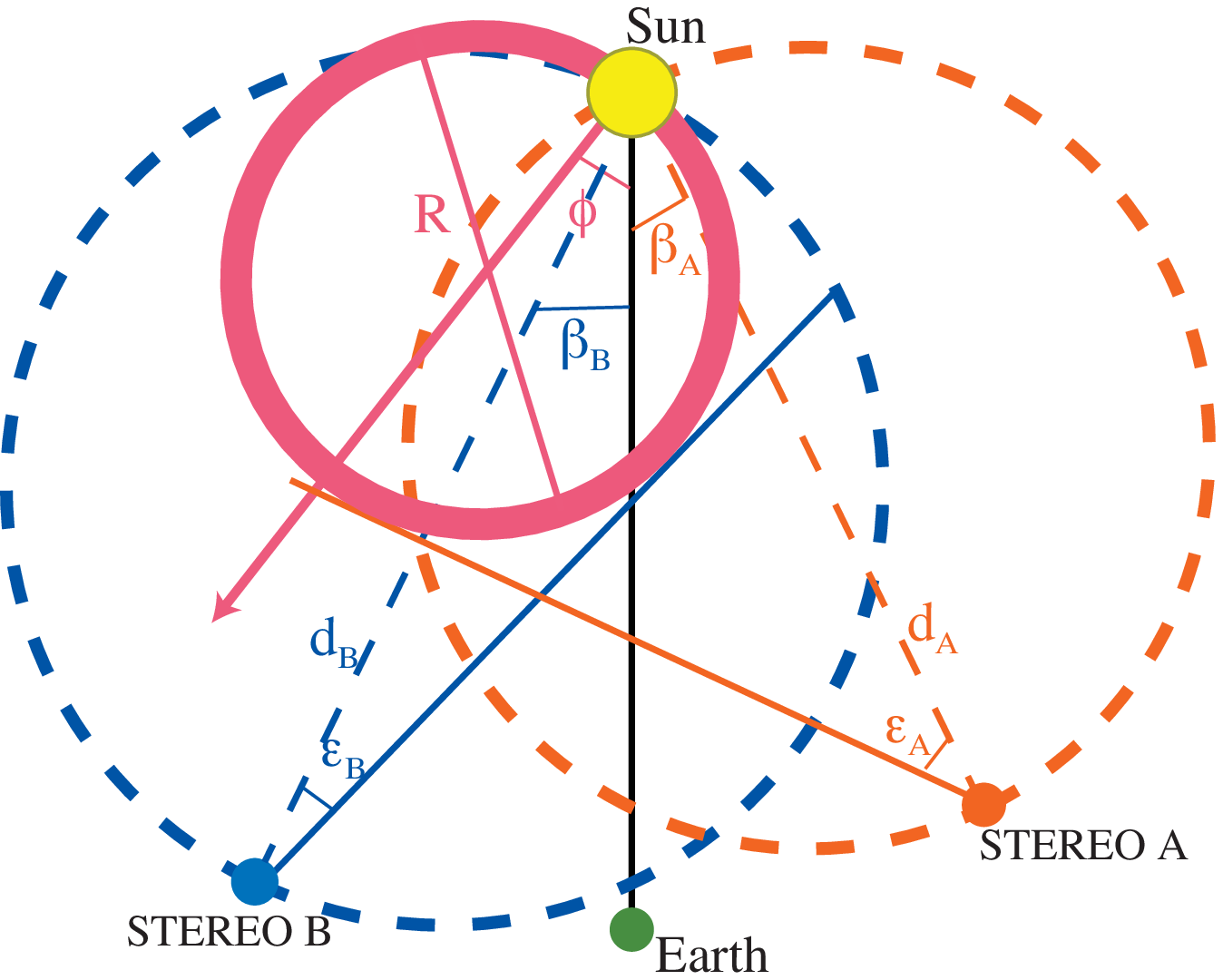}} \hspace{0.1cm}
{\includegraphics*[width=8.cm]{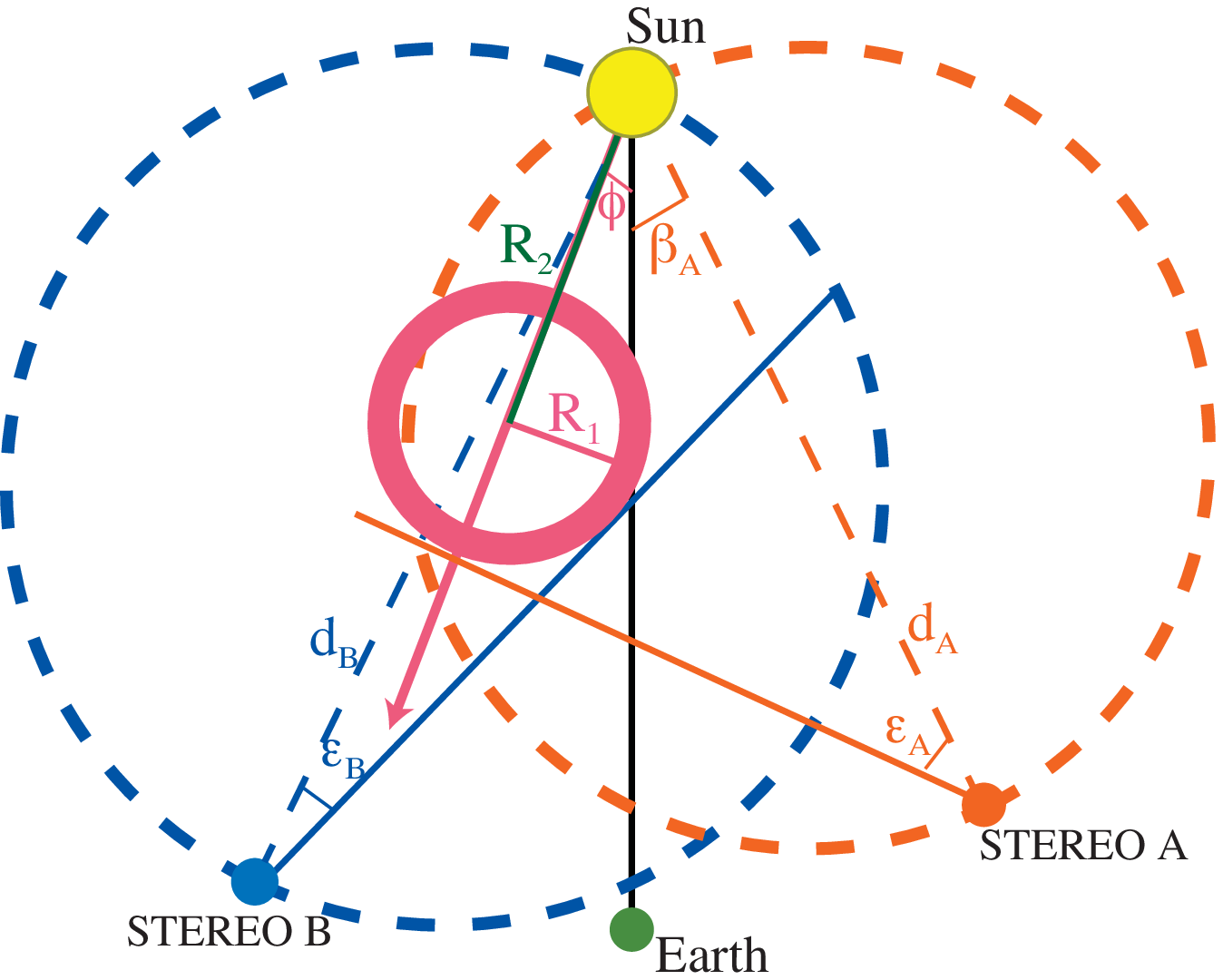}}
\caption{Sketches of the the two models used and described in this work, illustrating two ways to explain asymmetric observations. On the left, the CME is modeled as a sphere connected to the Sun with a varying direction of propagation $\phi$. On the right, the CME is modeled as a sphere on a fixed direction of propagation $\phi = -21^\circ$ but with a radius, $R_1$ varying with time. The sketches are to scale except for the size of the Sun, the Earth and the spacecraft and correspond approximatively to the geometry on April 26, 2008. The angles are $\epsilon_A = 39^\circ$ and $\epsilon_B = 18^\circ $. The varying $\phi$ model is with $\phi = -39^\circ$ and $R = 135~R_\odot$ and the varying radius model with $R_1 = 35~R_\odot$ and $R_2 = 90~R_\odot$.}
\end{center}
\end{figure*}

In this article, we discuss how simultaneous observations of a CME from the two {\it STEREO} spacecraft can be used to derive the CME central longitude or its radius of curvature, in addition to its radial distance. To do so, we propose two simple models of the CME described in section~\ref{model}. Contrary to triangulation techniques, we do not assume that the HIs are able to track the exact same feature for both spacecraft, but that they observe different parts of the same structure. In the first model, the CME cross-section is treated as an expanding-propagating circle attached to the Sun as proposed separately in \citet{Webb:2009}, \citet{Tappin:2009} and \citet{Lugaz:2009b}, but with a varying central position. In the second model, the CME has a fixed direction but it is not attached to the Sun and its diameter can vary freely. We apply these two models to the analysis of 4 CMEs observed in 2008 in section~\ref{data}.
The conclusions of this investigation are drawn in section \ref{conclusions}.

\section{DERIVATION OF CME CENTRAL LONGITUDE AND RADIUS FROM MULTIPOINT OBSERVATIONS} \label{model}
\subsection{Model 1: CME Central Longitude}\label{Phi}
CMEs propagating between the {\it STEREO-A} and {\it B} spacecraft, i.e. CMEs propagating approximatively towards Earth, can be imaged to large elongation angles by the HIs onboard both {\it STEREO} spacecraft. 
Taking into account the difference in spacecraft heliocentric distances ($d_A \sim$ 0.95~AU, $d_B \sim$ 1.05~AU), the fact that the HIs observe the same CME at the same time at different elongation angles can give us information about the CME central position. Such information cannot be deduced using the Point-P approximation, because it assumes a too simplistic CME geometry. Direct triangulation can be done \citep[]{Liu:2010}, but only under the assumption that both {\it STEREO} spacecraft observe the exact same plasma element, which is unlikely to be the case for most CMEs at large elongation angles. For some CMEs, direct triangulation might not give realistic results. We suspect that it is the case for the April 26, 2008 CME, because it appears as an halo CME in {\it STEREO-B}. The best fit of the method of \citet{Rouillard:2008} for this CME gives $\phi_A = -33.5 \pm18.0^\circ$ and $\phi_B = 2.1 \pm6.5^\circ$ for STEREO-A and B respectively; these two results are not consistent with each other.
However, one can use the expanding-propagating bubble approximation of \citet{Tappin:2009} and \citet{Lugaz:2009b} with two parameters to analyze the data from the two spacecraft simultaneously; for each pair of position angles (for example 90 and 270), the derived parameters are the diameter of the circular front and its direction of propagation (see left panel of Figure~1). 
The main assumptions of this model are that the CME cross-section is circular and that the CME is anchored at the Sun.

Assuming simultaneous observations by {\it STEREO-A} and {\it STEREO-B}, one can write an expression for the diameter of the sphere for each of the spacecraft following \citet{Lugaz:2009b}:
\begin{equation}
R_A  =  2 d_A \frac{\sin(\epsilon_A)}{1 + \sin(\beta_A - \phi + \epsilon_A)},
\end{equation}
where $\phi$ is the direction of propagation of the CME away from the Sun-Earth line (defined positive from B to A) and  $\beta_A$ is the angular separation (defined as a positive number) of the spacecraft with the Sun-Earth line. The same relation applies for spacecraft B by replacing $\phi$ by $-\phi$.
Solving for $R_A = R_B$ and regrouping the terms in $\phi$ gives:
\begin{equation}
\phi = \arcsin \left(\frac{P-1}{Q}\right) + \alpha 
\end{equation}
with
\begin{eqnarray*}
P & = & d_B \sin \epsilon_B / (d_A\sin \epsilon_A),\\
Q & = & \sqrt{P^2 + 2P \cos(\beta_B + \beta_A + \epsilon_B + \epsilon_A) +1}, \mathrm{and}\\
\tan \alpha & = & \frac{P \sin(\beta_A + \epsilon_A) - \sin(\beta_B + \epsilon_B)}{P \cos(\beta_A + \epsilon_A) + \cos(\beta_B + \epsilon_B)}.\\
\end{eqnarray*}
The diameter of the sphere can then be simply derived from equation (1). 

\subsection{Model 2: CME Radius}\label{Flat}
Another totally independent way to explain the different elongation measurements of the two spacecraft is to consider that it is due to the expansion of the CME front (see right panel of Figure~1). In this model, the CME is not anchored at the Sun and its radius is one of the derived parameters. To compute the CME radius, we have to consider the direction of propagation, $\phi$, as a known quantity. It can be, for example, obtained from the COR-2 fitting done by \citet{Thernisien:2009}. The main assumptions of this model are the circular cross-section and the absence of heliospheric deflection.

\begin{figure*}[ht*]
\begin{center}
{\includegraphics*[width=6cm]{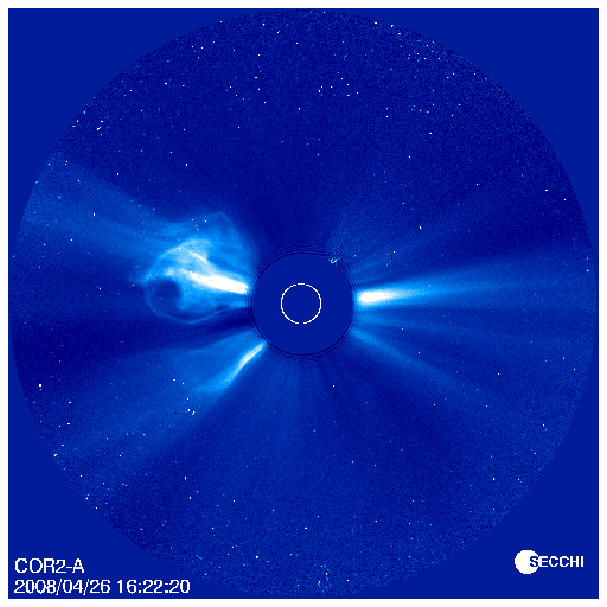}}
{\includegraphics*[width=6cm]{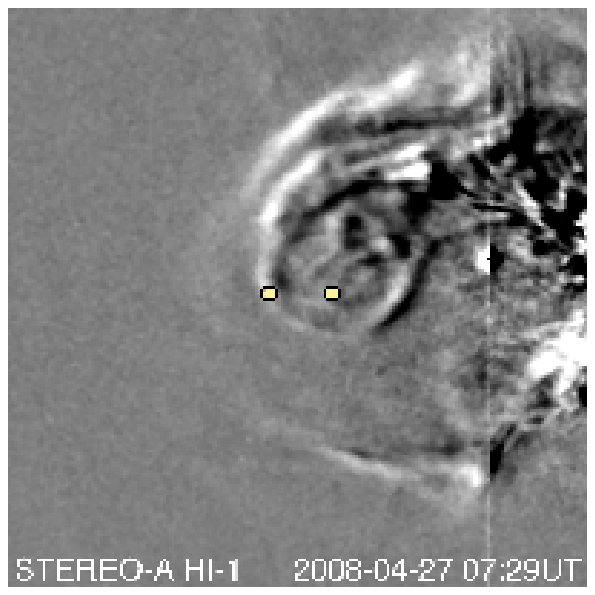}}
{\includegraphics*[width=6cm]{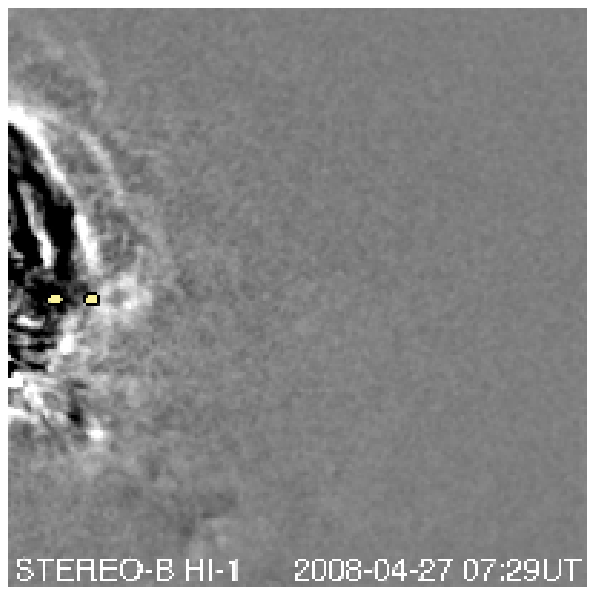}}
\caption{April 26, 2008 CME as observed by {\it SECCHI-A}/COR-2 ({\it left}) and HI-1 on {\it STEREO-A} ({\it midde}) and {\it STEREO-B} ({\it right}). The HIs images are taken 33 seconds apart, illustrating the asymmetric nature of the observations. The yellow dots illustrate the positions tracked by the NRL measurements (top and center of the ejecta).}
\end{center}
\end{figure*}

Noting $R_1$ as the radius of the CME, $R_2$ as the heliocentric distance of the center of the CME, $R_C$ as the distance of the center of the CME to the observer, and $\epsilon_A - \gamma$ as the angle between the Sun, STEREO-A and the center of the CME,  the law of cosines and the law of sines give, respectively:
\begin{eqnarray*}
R_C^2 & = & d_A^2 + R_2^2 - 2 d_A R_2 \cos(\beta_A - \phi)\\
R_2 \sin(\beta_A - \phi) &= & R_C \sin(\epsilon_A - \gamma)
\end{eqnarray*}
But $R_C \sin\gamma = R_1$ and $R_C \cos \gamma = \sqrt{R_C^2-R_1^2}$ and the second equation can be rewritten as:
$$
R_2 \sin(\beta_A - \phi) = \sqrt{R_C^2 - R_1^2} \sin \epsilon_A - R_1 \cos \epsilon_A
$$
which can be combined with the first equation into the following simple equation:
\begin{eqnarray}
& & R_2 \sin(\beta_A - \phi)  + R_1\cos\epsilon_A  \nonumber \\
&=& \sin\epsilon_A \sqrt{d_A^2 + R_2^2 - R_1^2 - 2d_A R_2 \cos(\beta_A -\phi) }, 
\end{eqnarray}
with the equation for spacecraft B obtained by replacing $\phi$ by $-\phi$.
By solving equations (3) for {\it STEREO-A} and {\it STEREO-B} simultaneously, $R_1$ and $R_2$ can be uniquely derived. The CME half-angle is given by $\theta = \arctan(R_1/R_2)$. It should be noted that for CMEs propagating directly towards Earth ($\phi \sim 0$), this model always predicts that the HI instruments onboard the two {\it STEREO} spacecraft will observe the exact same time-elongation profile, which is a direct consequence of the assumption of a circular front. Also, if $\epsilon_A < \epsilon_B$ at one time but $\epsilon_B < \epsilon_A$ at another time, there is no solution with $R_1 > 0$ and $R_2 > 0$, and the model cannot be applied. In these cases, the CME deformation, deflection and the deviation from circular of its cross-section must be taken into account. 

The main hypothesis of these two models is that the HI instruments onboard {\it STEREO-A} and {\it B} do not observe the same plasma element, as is likely to be the case in the COR fields-of-view and as is assumed in \citet{Liu:2010}. The assumption here is that the CME front is locally circular and that it is ``projected'' onto the HI fields-of-view at the elongation angle corresponding to the tangent to this front. 
Similar assumptions have been made by \citet{Wood:2009b} and \citet{Tappin:2009} to derive CME positions from observations at large elongation angles from one spacecraft.

\section{DATA ANALYSIS} \label{data}
\subsection{Data Selection}

For these new methods to be applied, the studied CMEs have to propagate between the two {\it STEREO} spacecraft, and they may appear as halo, partial halo or wide CMEs in LASCO field-of-view. To find potential CMEs to study, we look at all instances of CMEs with an apparent width larger than 100$^\circ$ in LASCO field-of-view from January  2008 when the spacecraft separation reached 45$^\circ$ until May 2009. We found 17 CMEs, 5 of which were observed simultaneously by the HI instruments onboard {\it STEREO-A} and {\it B}, they are the April 26, June 2, August 30, November 3 and December 12, 2008 CMEs. The November 3 is associated with an instance of CME-CME interaction and we exclude it from the current study \citep[]{Kilpua:2009}. Although the August 30 CME also interacts with a preceding ejection, the preceding ejection appears to be small enough in size so that it did not influence too strongly the propagation of the overtaking CME and, therefore, we keep this CME in this study. The December 12 CME had two bright fronts observed in both spacecraft and we analyzed 3 different datasets for the April 26 CME, resulting in a total of 7 analyzed pairs of datasets. We first apply the two models to a detailed case study for the April 26, 2008 CME, before presenting the results for the other CMEs.

\begin{figure*}[ht]
\begin{center}
{\includegraphics*[height=5cm]{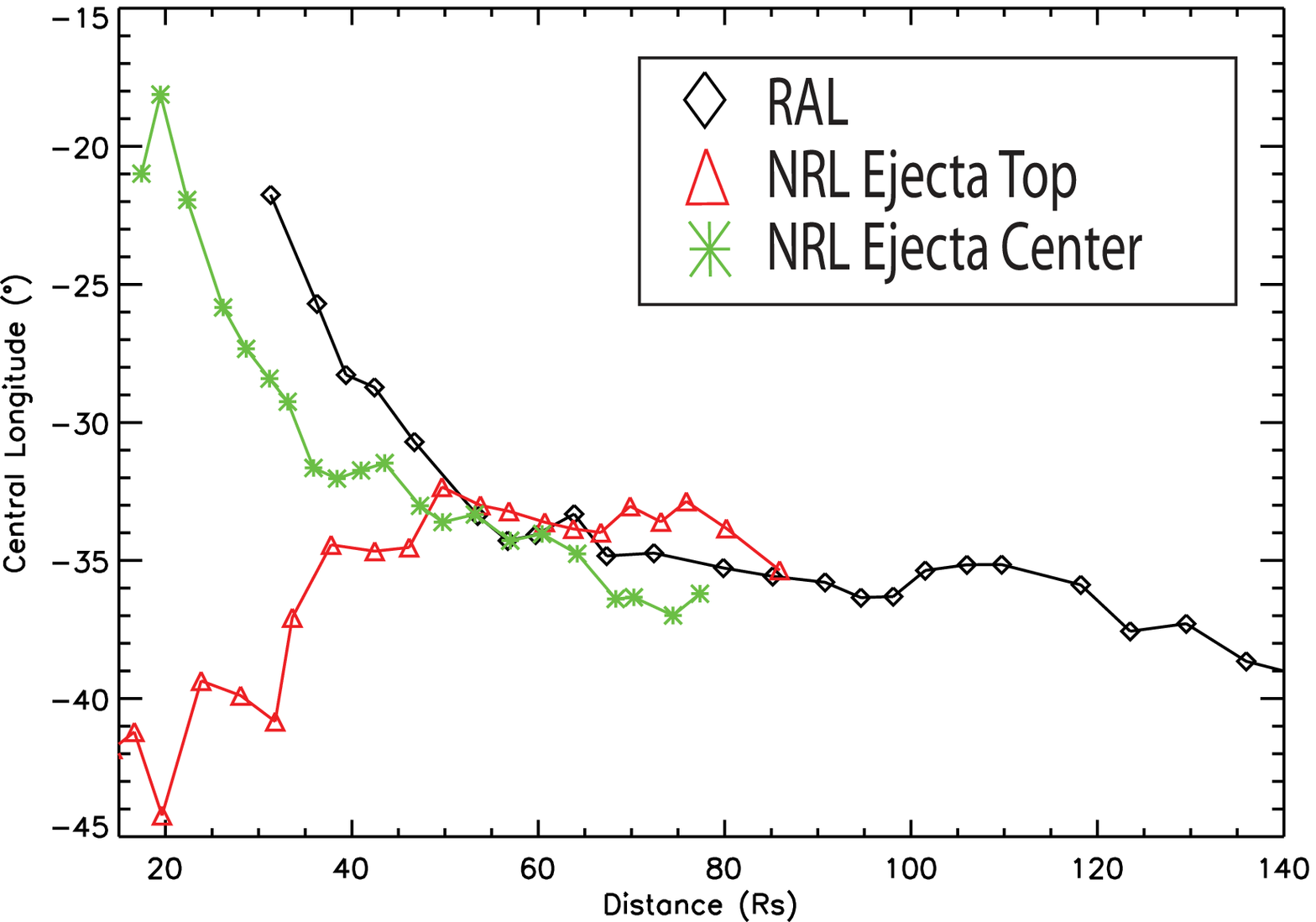}}
{\includegraphics*[height=5cm]{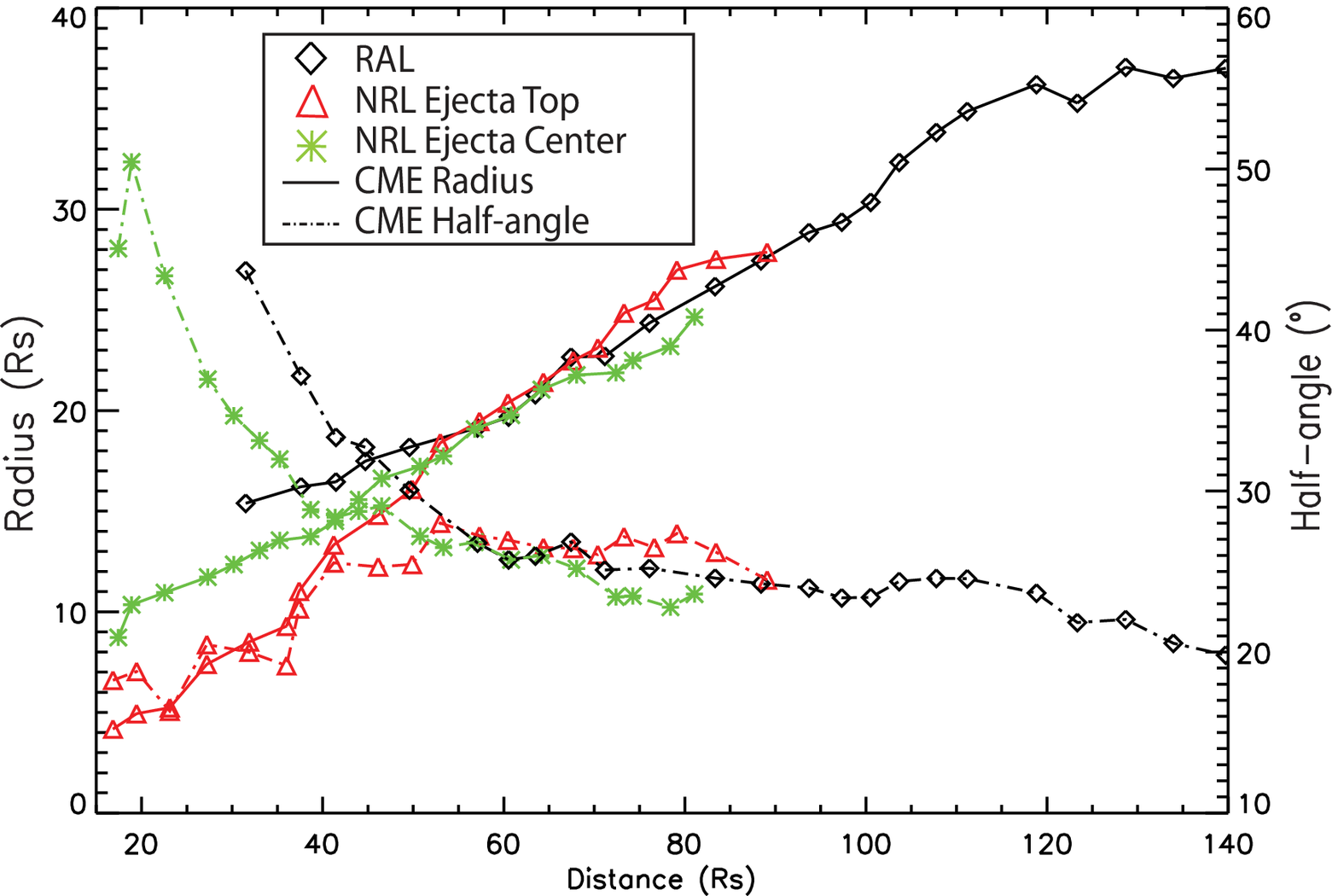}}\\
{\includegraphics*[height=5cm]{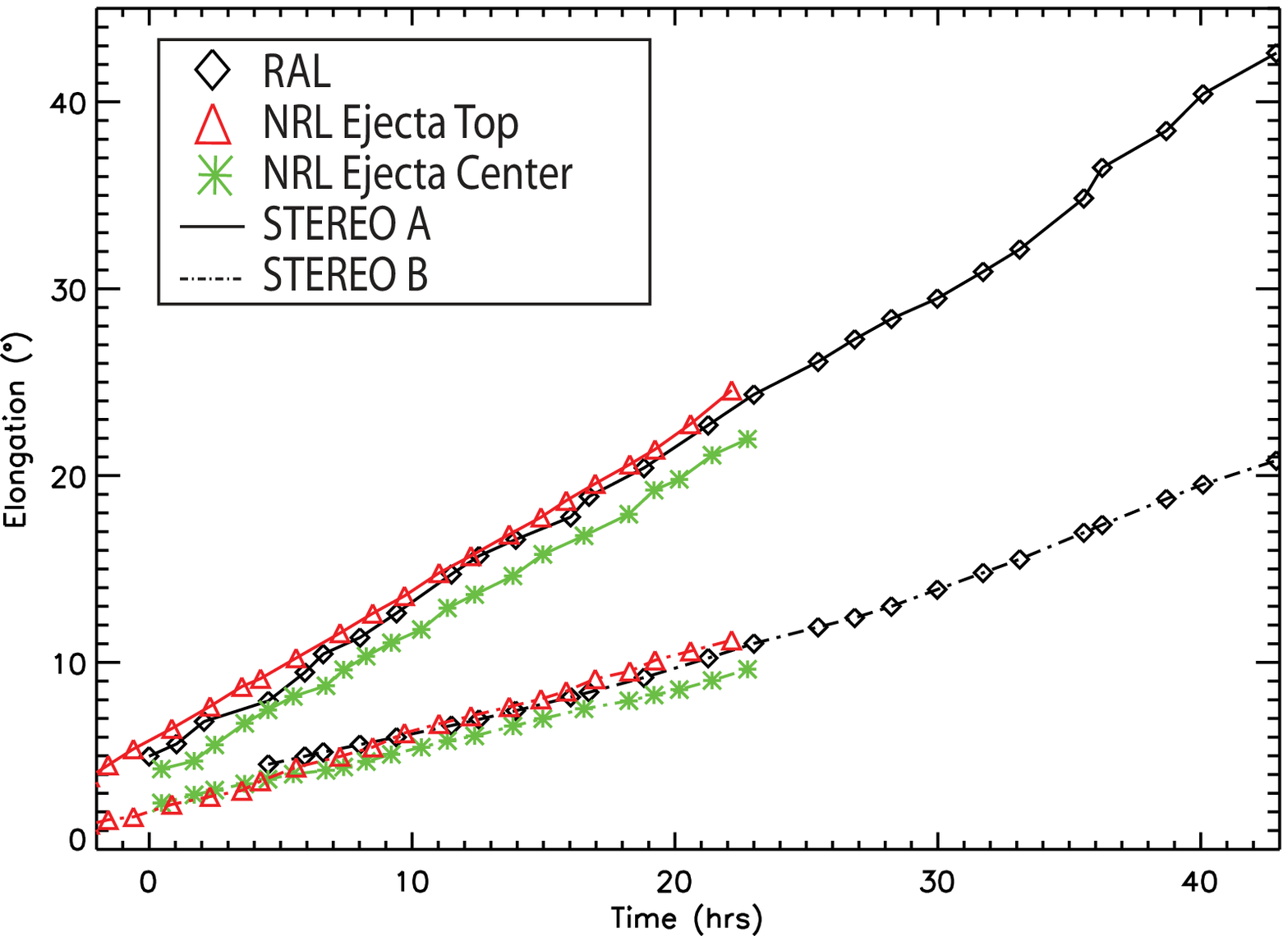}}
{\includegraphics*[height=5cm]{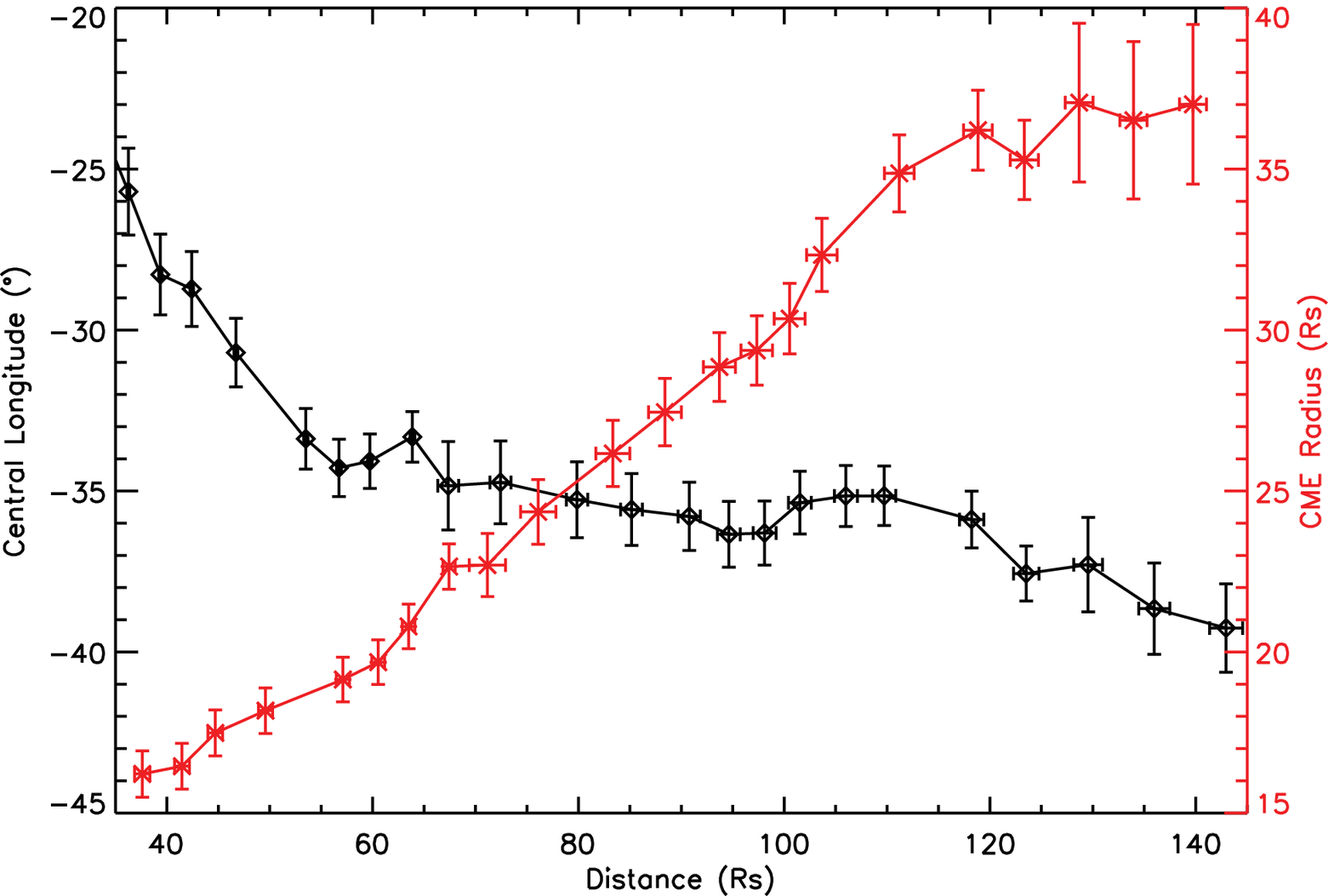}}
\caption{{\it Top Left}: Central longitude, $\phi$, as derived from the varying-$\phi$ model (model 1). {\it Top Right}: CME radius, $R_1$ (solid), and half-angle, $\theta$ (dash-dot), as derived from the varying-radius model (model 2). {\it Bottom Left}: Elongation measurements of the CME, showing the RAL measurement and the NRL measurements for the top and the center of the CME. {\it Bottom Right}: Error for the RAL measurements assuming the elongation angles are measured with a precision of 15 pixels.}
\end{center}
\end{figure*}

\subsection{April 26 CME}

On April 26, 2008 at 1425UT, SECCHI-A/COR-1 observed an eastern limb CME. This event was studied by \citet{Thernisien:2009} and \citet{Colaninno:2009} in the COR-1 and COR-2 fields-of-view and by \citet{Wood:2009} in the entire SECCHI field-of-view. 
Images of the CME in COR2-A, HI1-A and HI1-B are shown in Figure~2. To do our analysis, we used elongation measurements for PA 90 and 270 obtained from the Naval Research Laboratory (hereafter, NRL measurements) from J-maps including COR and HI observations and from the Rutherford Appleton Laboratory (hereafter, RAL measurements) from J-maps derived from HI-only  observations along the ecliptic. The time-elongation plots are shown on the bottom left panel of Figure~3. The tracked front corresponds to the top of the ejecta for the RAL and the first of the NRL measurements and to the center of the ejecta for a second set of NRL measurements. 
Two difficulties with applying the models described above to real data are that the elongation measurements must be derived accurately and that the boundary of the same structure must be tracked in {\it STEREO-A} as well as in {\it STEREO-B}. This last condition may not always be fulfilled in HI-2 where the cause of bright fronts is sometimes hard to establish \citep[e.g., see][]{Lugaz:2008}.

We analyzed the time-elongation data for {\it STEREO-A} and {\it B} for this CME with the two methods proposed above. Using data from two different groups allow us to quantify the errors associated with the manual determination of the elongation angles. The results of our analysis are shown in Figure~3. In the bottom right panel of this Figure, we show the error bars for model 1 (black) and model 2 (red) assuming the elongation measurements are made with a precision of 15 pixels corresponding to an uncertainty of $\pm .15^\circ$ and $\pm .5^\circ$ in HI-1 and HI-2 fields-of-view, respectively. The error in direction (model 1) is typically $\pm 2^\circ$ for both instruments and the error in the CME radius (model 2) is $\pm 1~R_\odot$ in HI-1 and increases to $\pm 2.5~R_\odot$ when the CME is observed in both HI-2 simultaneously.

According to model 1, the CME is deflected towards the East (i.e. away from Earth) reaching a near-constant central longitude of -35$\pm 2^\circ$ at about 50~$R_\odot$, the CME continues to be deflected towards the East with a rate of about -3.5$^\circ$/day from this distance on (see top left panel of Figure~3). The initial angle is $\sim -20^\circ$ comparable to the angle of -21$^\circ$ derived  by \citet{Thernisien:2009} based on COR2 data. 
The central position of the CME using only {\it STEREO-A} data and the Fixed-$\phi$ procedure of \citet{Rouillard:2008} can be estimated at -33.5$\pm 18^\circ$, while \citet{Wood:2009} reports a best-fit value of -26.5$^\circ$. Our value of -$35 \pm 2^\circ$ appears in relatively good agreement with these values. 

For model 2, we assume the fixed direction of propagation $\phi = -21^\circ$ from \citet{Thernisien:2009}. Then, the CME radius monotonically increases from a value of 8$\pm 3~R_\odot$ at 20~$R_\odot$ to 26$\pm 1~R_\odot$ at 80~$R_\odot$, and to about 37~$R_\odot$ at 140~$R_\odot$. The corresponding CME half-angle is equal to $25 \pm 2^\circ$ from 60~$R_\odot$ to 120~$R_\odot$ with a general rate of shrinking of 4$^\circ$/day until the end of the measurements (see top right panel of Figure~3). 

\begin{figure*}[ht]
\begin{center}
{\includegraphics*[width=8cm]{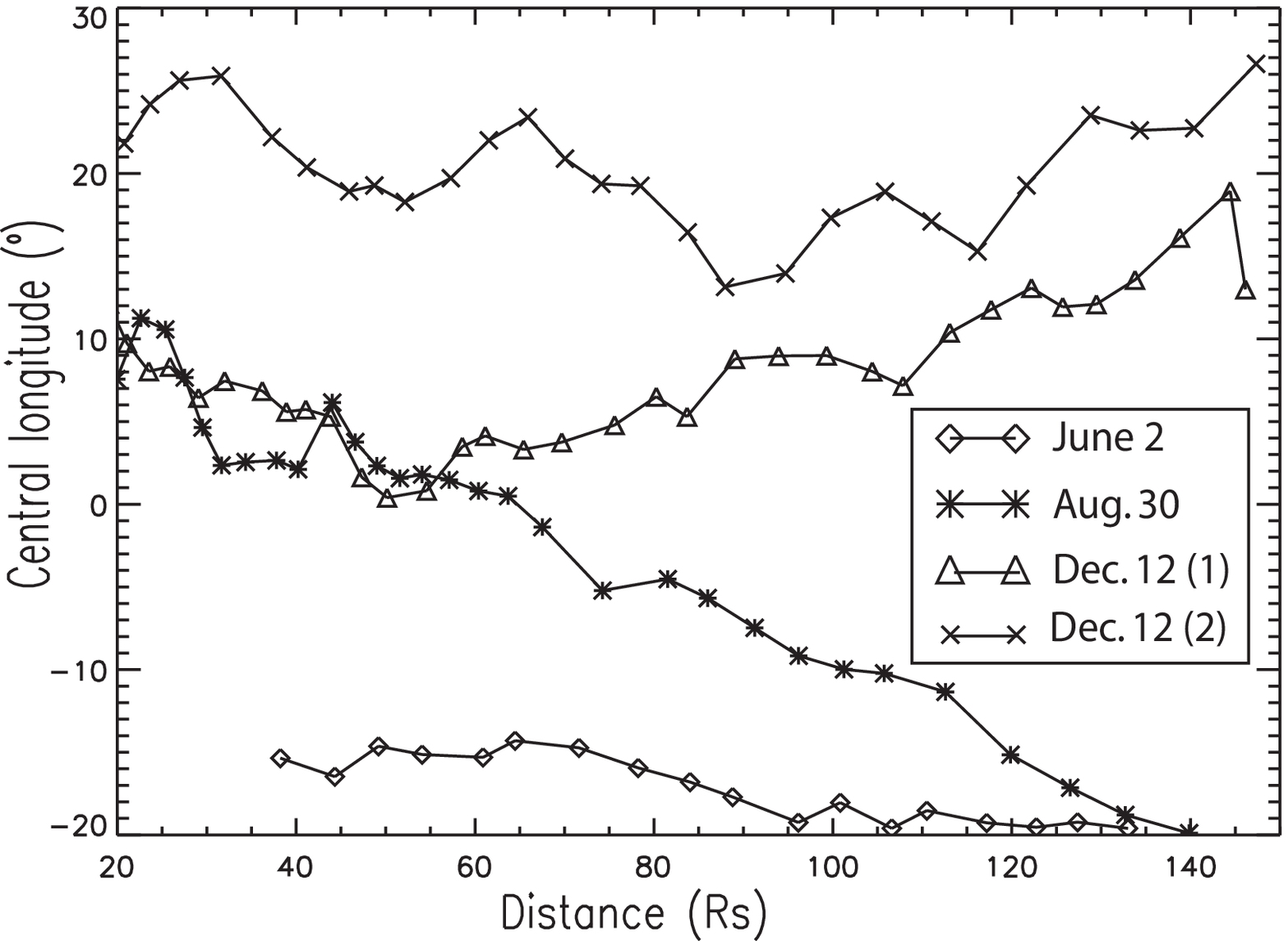}}
{\includegraphics*[width=8cm]{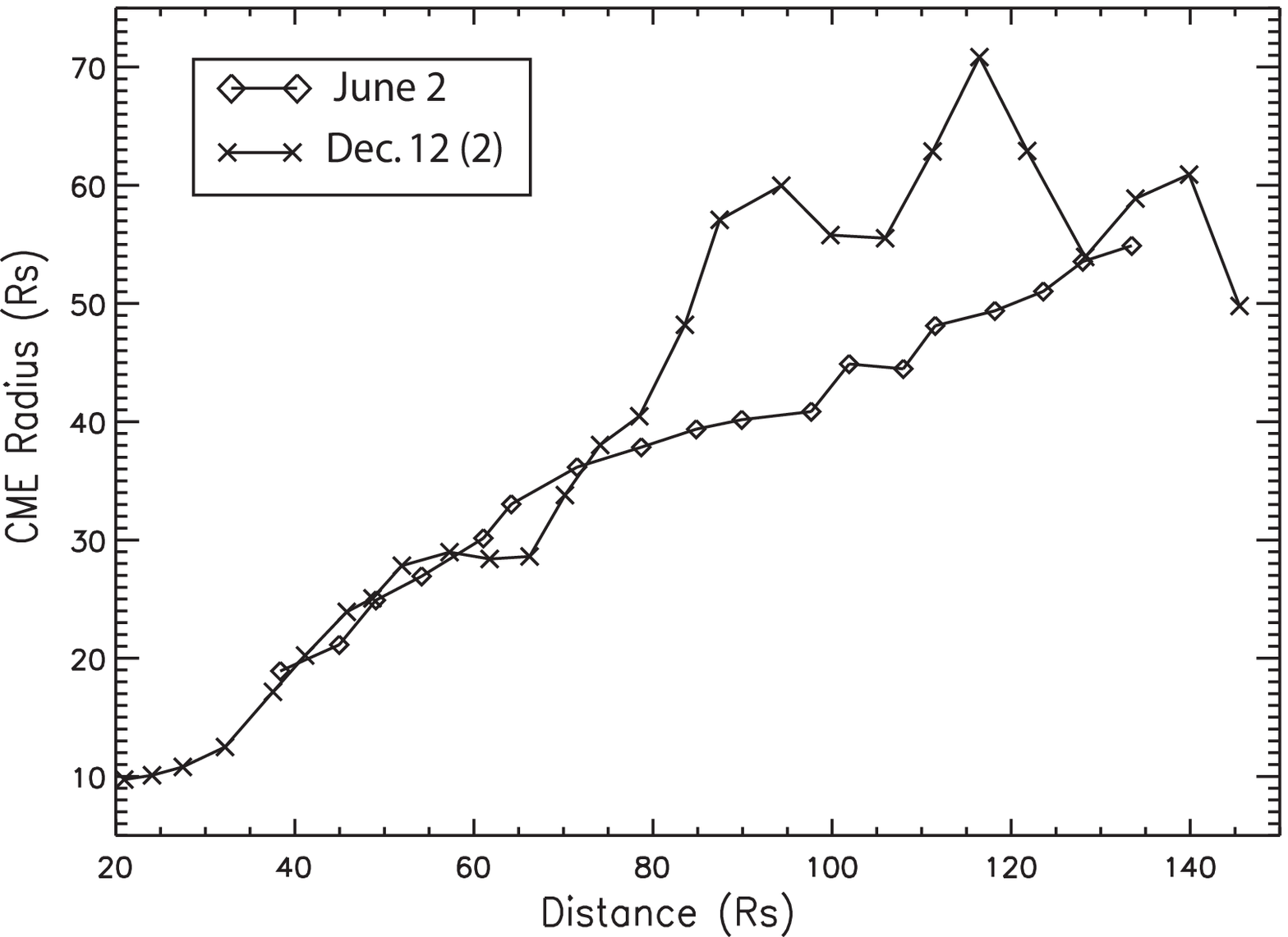}}
\caption{{\it Left}: Central longitude, $\phi$, as derived with the varying-$\phi$ model (model 1) for the June 2 and August 30 CMEs and for the two bright features associated with the December 12 CME. {\it Right}: CME radius, $R_1$, as calculated for model 2 for the June 2 CME and the second front of the December 12 CME.}
\end{center}
\end{figure*}

In the first 40$~R_\odot$, the results using the different datasets are inconsistent with each other (see top panels of Figure~3). This is because the CME appears in {\it STEREO-B} as an halo and in {\it STEREO-A} as a limb CME, and we are not able to identify the same front in the observations from the two spacecraft. In fact, the CME is first visible in HI-B  (at $\sim 4^\circ$) when it reaches a distance of about $32~R_\odot$ from the Sun, corresponding to observations around 8$^\circ$ elongation in HI-A (see bottom left panel of Figure~3).  Before this time, since there is no measurement of the CME in HI-B, we use the elongation angles measured by {\it STEREO-B} in COR-2 for the NRL data. We find it impossible to separate the top of the ejecta from the shock front and piled-up mass in COR2-B and early on in HI1-B. Therefore, we are not able to identify the same front between {\it STEREO-A} and {\it STEREO-B} when the CME is within 40~$R_\odot$ from the Sun, and the results are inconsistent between the RAL and NRL measurements. For the center of the ejecta, we are able to identify the common feature in both {\it STEREO-A} and {\it B} and the models, in this case, give a more realistic evolution of $\phi$ and $R_1$. This is because in {\it STEREO-B}, where the CME appears as an halo, it is still relatively easy to separate the center of the ejecta from the other density features. It should be noted that the center of the ejecta might be of relatively small angular extent in the azimuthal direction, in which case, using direct triangulation as in \citet{Liu:2010} is more adapted. 

We compared the CME distance as derived with these two models, with the HM model proposed in \citet{Lugaz:2009b} using only  {\it STEREO-A} data. All methods yield the same distances for the CME within 10$\%$. However, from the two new models, it is possible to predict a hit/miss at different spacecraft's positions in the heliosphere based on the azimuthal properties of the CME. Based on the height-time profile in the heliosphere using the second model, we fit the data to a CME with a constant speed of 534~km~s$^{-1}$ --to compare to the final speed of 543~km~s$^{-1}$ derived by \citet{Wood:2009}--. Additionally, we calculate the CME half-angle, $\theta$, with the fixed rate of $-4^\circ$/day, using the best-fit formula of $\theta = 25^\circ - 4^\circ \times (t - t_0)$, where $t$ is the time in day since $t_0$ = April 27 1200UT. With these parameters, the model correctly predict that the CME does not hit ACE but it predict that the CME hit {\it STEREO-B} at 0300UT on 04/30. In-situ observations by ACE and the two {\it STEREO} spacecraft show that only {\it STEREO-B} detected the passage of an iCME from 1530UT on 04/29 to 0700UT on 04/30, which translates into an error of about 11 hours for the arrival time for our model.

\subsection{June 2, August 30 and December 12 CMEs}

The June 2, 2008 CME was included in the study by \citet{Thernisien:2009} and studied in \citet{Robbrecht:2009}, while the December 12, 2008 CME has been analyzed by \citet{Davis:2009} and \citet{Liu:2010}. For our study, we used the data available from the Rutherford Appleton Laboratory website. 
In the left panel of Figure~4, we show the variation of the central longitude of the CMEs following model 1 for the 4 datasets. In Table~1, we compare our results with information available from the flare location, with the direction of propagation as calculated with the procedure of \citet{Rouillard:2008} and with other published analyses. It should be noted that the method of \citet{Liu:2010} assumes that $d_A = d_B$, whereas we use the real values of the spacecraft heliocentric distances (which we assume constant over the duration of an event). In our experience, doing so is required to obtain consistent values, especially at large elongation angles. For example, analyzing the last pair of elongation angles from our measurements for the second front of the December 12 event, the method of \citet{Liu:2010} gives $-5^\circ$ with $d_A = d_B$, but using the exact values of  $d_A$ and $d_B$ shifts the derived longitude from -5$^\circ$ to 13$^\circ$. This corrected value is closer to the last value of 27$^\circ$ obtained with model 1 than the value of $-5^\circ$ reported by \citet{Liu:2010}.

\begin{table*}[ht]
\begin{center}
\begin{tabular}{|c|c|c|c|}
\tableline
CME & Method \& Instruments (Reference) & Phi Estimate\\
\tableline
Apr. 26 & Flare  & -9$^\circ$\\
Apr. 26 & Visual COR2s (1) & -21$^\circ$ \\
Apr. 26 & Visual STEREOs (2) & -28$^\circ$\\
Apr. 26 & Mass COR2s (3) & -48$^\circ$  \\
Apr. 26 & Fixed-$\phi$ HI-A& -33.5 $\pm~18^\circ$ \\
Apr. 26 & Fixed-$\phi$ HI-B&  2.1 $\pm~6.5^\circ$\\
Apr. 26 & Model 1 & -17$^\circ$ (at 20~$R_\odot$) to -37.5$^\circ$ (at 130~$R_\odot$)\\
\tableline
June 2 & Visual COR2s (1) & -37$^\circ$ \\
June 2 & Fixed-$\phi$ HI-A & -24.2 $\pm~5^\circ$ \\
June 2 & Fixed-$\phi$ HI-B &  20.9 $\pm~11.5^\circ$\\
June 2 & Model 1 & -17 $\pm 2.7^\circ$\\
\tableline
Aug. 30 & Fixed-$\phi$ HI-A  & -16 $\pm~17^\circ$ \\
Aug. 30 & Fixed-$\phi$ HI-B & 19.2 $\pm~10.5^\circ$\\
Aug. 30 & Model 1 & 10 $^\circ$ (at 20~$R_\odot$) to -20$^\circ$ (at 140~$R_\odot$)\\
\tableline
Dec. 12 Front 1& Fixed-$\phi$ HI-A & -14.7 $\pm~13^\circ$\\
Dec. 12 Front 1& Fixed-$\phi$ HI-B &  12.6 $\pm~6.5^\circ$\\
Dec. 12 Front 1 & Triangulation (4) & 0 $\pm~5^\circ$\\
Dec. 12 Front 1& Model 1 & 10 $\pm~10^\circ$\\
\tableline
Dec. 12 Front 2 & Fixed-$\phi$ HI-A & 8.3 $\pm~4.5^\circ$ \\
Dec. 12 Front 2 & Fixed-$\phi$ HI-B & -1.5 $\pm~7^\circ$ \\
Dec. 12 Front 2 & Triangulation (4) & 5 $\pm~3^\circ$ (up to 70~$R_\odot$) -3 $\pm 5^\circ$ (after) \\ 
Dec. 12 Front 2 & Model 1 &  20 $\pm~7^\circ$ \\
\tableline
\end{tabular}
\caption{Direction of propagation for the four CMEs from this work compared to other methods. References: 1 Thernisien et al. (2009), 2 Wood \& Howard (2009), 3 Colaninno \& Vourlidas (2009), 4 Liu et al. (2010).}
\end{center}
\end{table*}

We find that the June 2 CME and the two fronts from the December 12 CME propagate close to radially outward until about 140~$R_\odot$ (0.65~AU), while the August 30 CME shows a deflection towards the East of about 30$^\circ$ during its propagation and crosses the Sun-Earth line. As noted in section~2.2, the second model, because it assumes a fixed direction of propagation cannot be used for CMEs propagating exactly towards Earth ($\phi \sim 0^\circ$) nor for CMEs which cross the Sun-Earth line during their propagation, therefore we only applied this model with the measurements from the June 2 CME as well as the second front of the December 12 CME, which appear to propagate away from the Sun-Earth line. We used directions of propagation of -15$^\circ$and 20$^\circ$ for these features, respectively. From model~1, the June 2 CME appears to move on a  quasi-radial trajectory at about $-15^\circ$, while the second front of the December 12 CME appears to propagate on a direction of about 20$^\circ$ with respect of the Sun-Earth line; hence, we choose these numbers for the fixed directions of propagation. The results of the second model are shown in the right panel of Figure~4. 

The June 2 observations can be explained, using model 2, by a CME which propagates on a fixed radial trajectory and whose radius in the ecliptic increases more slowly than self-similarly, the CME half-angle decreasing from $45^\circ$ to $35^\circ$ (see Figure~4). Alternatively, it can be explained by model 1 as a CME whose direction of propagation is deflected by about 5$^\circ$ towards the East in about 0.45~AU. Both these explanations appear physically realistic, involving a propagation close to radially outward and an evolution close to self-similar, and it is likely that the evolution of the June 2 CME is a combination of these two results, with a limited eastward deflection and a small shrinking of the CME front. The observations from the August 30 CME cannot be explained by model 2, because the CME appears first farthest in {\it STEREO-B} before appearing farthest in {\it A} (as noted in 2.2, this causes model 2 to be inapplicable). Using model 1, it corresponds to a CME being deflected by 30$^\circ$ towards the East in 0.5 AU. The 2 features from the December 12 CME appear to propagate close to the Sun-Earth line on near-radial trajectories but they cannot be simply analyzed with model 2, either because the fronts propagate too close from the Sun-Earth line or because the model's assumption of observing the tangents to a circular CME cross-section is not correct for this CME. It appears from both our study and that by \citet{Liu:2010} that the first front of this CME propagates about 5-10$^\circ$ East of the second front.

\section{DISCUSSION AND CONCLUSIONS}\label{conclusions}
In this article, we propose two models to derive information about the azimuthal properties of CMEs from multi-spacecraft observations in the heliosphere using simple geometrical considerations. These models can be used to derive the CME radius or central position from SECCHI observations without having to rely on any extra human judgement or fitting procedure, after obtaining the time-elongation measurements. The main hypothesis of these two models is that the HI instruments onboard {\it STEREO-A} and {\it B} do not observe the same plasma element, as is likely to be the case in the COR fields-of-view and as is assumed in \citet{Liu:2010}. The assumption here is that the CME front is locally circular and that it is viewed in the HIs at the elongation angle corresponding to the tangent to this front. We applied these two new models to six features belonging to four CMEs observed in 2008 by both STEREO spacecraft. 

For five of the six studied features, the April 26 (center and front), June 2 and December 12, 2008 (front 1 and front 2) CMEs, we find that the measurements can be explained as being from CMEs propagating close to radially outward. However, for two of the CMEs (the June 2 CME and both features from the December 12 CME), we do not assume a radial propagation and we derive the CME central longitude directly with model 1. We find that the CME central longitude remains within 15$^\circ$ of radial, with the June 2 CME, in particular, being deflected monotonically towards the east by about 5$^\circ$ in 0.5~AU. For these CMEs, model 2, which assumes radial propagation but does not assume a self-similar expansion, does not provide additional information. Also, the two features associated with the December 12 CME are found to be separated by about 10$^\circ$ at all time, which was also found using direct triangulation by \citet{Liu:2010}. 


For the April 26, 2008 CME, the model assuming self-similar expansion (model 1) appears to fail, because it predicts a large deflection of the CME which is not physically expected nor confirmed by in-situ measurements. Self-similar expansion of CMEs in the heliosphere has long been assumed \citep[]{Xue:2005, Krall:2006, Wood:2009}, but in general, it has not been tested with heliospheric observations. We analyze this CME with model 2 , where self-similar expansion is not assumed and using the direction of propagation derived by \citet{Thernisien:2009}.  We derive the change over time of the CME cross-section, which is found to decrease with a rate of about $4^\circ$/day. The cause of this decreasing cross-section has to be further studied, but it shows that the CME expands slower than self-similarly. This may, for example, reflect a variation of the CME radius of curvature in the ecliptic plane. This result is found for solar minimum conditions when the background solar wind is more simply structured.  For the other CME (August 30, 2008) the measurements can only be explained, with model 1, by a a deflection towards the East of $30^\circ$ in 0.5~AU of the CME. It is also possible that an instance of CME-CME interaction locally deformed this CME front away from circular which would render our models less accurate.


In this study, we ignore the effect of the Thomson sphere and the angular dependence of the Thomson scattering intensity. To apply the new analysis techniques presented here, the CME has to propagate between the {\it STEREO-A} and {\it STEREO-B} spacecraft. For this study focusing on CMEs observed in 2008, this means that the CMEs propagate within 40$^\circ$ from the Sun-Earth line, and we believe that ignoring the difference in Thomson scattering between the two spacecraft is justified as a first approximation. However, as the two {\it STEREO} spacecraft continue to separate, CMEs propagating farther away from the Sun-Earth line will be observed simultaneously by both spacecraft and a more thorough analysis of the effect of the two different Thomson spheres should be made. Additionally, effects associated with the interaction of a CME with the Thomson sphere have been found to happen at elongation angles greater than $45^\circ$ (e.g. \citet {Manchester:2008}, \citet{Lugaz:2008}), which is farther than that analyzed here.

We believe that, by providing a value of the CME radius in the ecliptic plane or the time-dependent variation of the central longitude which can be simply computed from time-elongation measurements,  these models could improve space weather forecasting of CMEs. However, to use such techniques on a real-world basis, it might be useful to have dedicated spacecraft making heliospheric observations at fixed locations, for example from the Lagrangian points L4 and L5. In general, we believe our model 1 can be used to derive the CME central longitude and its temporal variation but is not applicable for these CMEs whose expansion is not self-similar. When this is the case, for example because of interaction with the structured solar wind or with previous CMEs, we propose a second model, which can quantify the deviation from self-similar, assuming radial propagation. Future work should focus on the radius and central longitude of a CME at different position angles and on further testing and validation of the models. 

\begin{acknowledgments}
The research for this manuscript was supported by  NSF grant ATM0819653 and NASA grants NNX07AC13G and NNX08AQ16G.
J.~N.~H.-C.'s stay in Hawaii was supported by NSF grant ATM-0639335 and the NSF/REU program. 
We would like to thank Ying Liu and an anonymous reviewer for helping us improve this article.
SoHO and {\it STEREO} are projects of international cooperation between ESA and NASA. The SECCHI data are produced by an international consortium of 
  Naval Research Laboratory, Lockheed
  Martin Solar and Astrophysics Lab, and NASA Goddard Space Flight
  Center (USA), Rutherford Appleton Laboratory, and University of
  Birmingham (UK), Max-Planck-Institut f{\"u}r Sonnensystemforschung
  (Germany), Centre Spatiale de Liege (Belgium), Institut d'Optique
  Th{\'e}orique et Appliqu{\'e}e, and Institut d'Astrophysique
  Spatiale (France). 

\end{acknowledgments}

\end{document}